\documentstyle[12pt]{article}
\begin{document}
\newcommand{\rf}[1]{(\ref{#1})}
\newcommand{\beq}{\begin{equation}}
\newcommand{\eeq}{\end{equation}}
\newcommand{\bdm}{\begin{displaymath}}
\newcommand{\edm}{\end{displaymath}}
\newcommand{\bea}{\begin{eqnarray}}
\newcommand{\eea}{\end{eqnarray}}
\newcommand{\nn}{\nonumber \\}
\newcommand{\al}{\alpha}
\newcommand{\g}{\gamma}
\newcommand{\del}{\delta}
\newcommand{\Del}{\Delta}
\newcommand{\e}{\epsilon}
\newcommand{\ve}{\varepsilon}
\newcommand{\mn}{\mu\nu}
\newcommand{\om}{\omega}
\newcommand{\pa}{\partial}
\newcommand{\na}{\nabla}
\newcommand{\k}{\kappa}
\newcommand{\lam}{\lambda}
\newcommand{\vp}{\varphi}
\newcommand{\bfB}{\bf{B}}
\newcommand{\bfv}{\bf{v}}
\newcommand{\bfx}{\bf{x}}
\newcommand{\bfy}{\bf{y}}
\newcommand{\vx}{\vec{x}}
\newcommand{\vy}{\vec{y}}
\newcommand{\oB}{\overline{B}}
\newcommand{\oI}{\overline{I}}
\newcommand{\oR}{\overline{R}}
\newcommand{\cD}{\cal{D}}
\newcommand{\cL}{\cal{L}}
\newcommand{\cO}{{\cal{O}}}
\newcommand{\ra}{\rightarrow}
\newcommand{\ti}{\times}
\def\sepand{\rule{14cm}{0pt}\and}
\def\gtwid{\raise.3ex\hbox{$>$\kern-.75em\lower1ex\hbox{$\sim$}}}
\def\ltwid{\raise.3ex\hbox{$<$\kern-.75em\lower1ex\hbox{$\sim$}}}

\topmargin -0.8cm
\headheight 0pt
\headsep 0pt
\topskip -9mm

{\title{{\hfill {{\small NBI-HE-93-33\ \ \ \ \ \ }}\vskip 2cm}{\bf On
Primordial Magnetic Fields of Electroweak Origin }}


\author{
{\sc K. Enqvist} \\
{\sl Nordita} \\
{\sl Blegdamsvej 17} \\
{\sl DK-2100 Copenhagen} \\
{\sl Denmark}\\
{}~~\\
and\\
{}~~\\
{\sc P. Olesen }\\
{\sl The Niels Bohr Institute} \\
{\sl University of Copenhagen} \\
{\sl Blegdamsvej 17} \\
{\sl DK-2100 Copenhagen} \\
{\sl Denmark} \\
\sepand
}
\maketitle}
\vspace{3cm}
\begin{abstract}
We consider Vachaspati's primordial magnetic field which is generated at the
electroweak phase transition.
Assuming that either the gradients of the Higgs field  or,
alternatively,  the magnetic field itself are stochastic variables
with a normal distribution, we find that the resulting magnetic
field has an {\em rms} value in the present-day universe which is
fully consistent with what is required for the galactic dynamo mechanism.
\end{abstract}
\newpage
\section{Introduction}
The magnetic fields observed in galaxies (of the order
$10^{-6} $ G) can be understood as an amplification by a dynamo effect
\cite{dynamo} of a weak seed field of order $10^{-18} $ G on a
co-moving scale of 100 kpc.
The intergalactic  plasma has a large electrical conductivity, and the
magnetohydrodynamical equation
\beq
\pa {\bfB}/ \pa t = \na \ti ({\bfv} \ti {\bfB}) - \na \ti (\eta
\na \ti {\bfB})~~~~~,                                      \label{1}
\eeq
where $\eta$ is the inverse electrical conductivity, then implies
that the magnetic lines of force are essentially ``frozen into'' the fluid.
The magnetic flux through any contour moving with the plasma is thus
constant. The collapse of the plasma into  a galaxy
enhances the magnetic field (by a factor $\sim 10^4$), whereas
the remaining necessary enhancement is due to the differential
rotation and turbulent motion of the
plasma in the galaxy.

Eq. \rf{1} is homogeneous in $\bfB$, so if initially one has no field
it follows that a field can never be generated.
This is basically the reason for the need of a seed field.

It has often been speculated that the seed field is of primordial
origin, which means that it should be explained by features relevant
for the early universe.
Electromagnetism first occurs when the standard electroweak
$SU(2) \otimes U(1)_Y$ theory is broken down to $U(1)_{\rm em}$.
It is therefore particularly attractive that Vachaspati \cite{vach}
has explained the origin of a primordial field in terms of the
cosmological boundary condition that all physical quantities
should be uncorrelated over distances greater than the horizon
distance.
Since we start with the group $SU(2) \otimes U(1)_Y$ before the
electroweak phase transition, the resulting electromagnetic field
can be constructed in a way which is different from the usual
$F_{\mn} = \pa_\mu A_\nu - \pa_\nu A_\mu$.
The result is \cite{vach}
\bea
F_{ij} & = & - i(V^\dagger_i V_j - V_i V^\dagger_j)~~, \nn
V_i & = & \frac{2}{|\phi|} \sqrt{\frac{\sin \theta}{g}}~
\pa_i \phi ~~,                                           \label{2}
\eea
where $\phi$ is the Higgs field.
At the electroweak phase transition the correlation length in the
broken phase is
$\sim 1/m_W$ (assuming that the Higgs mass is comparable to
$m_W$).
The field $F_{ij}$ is thus constant over a distance $\sim 1/m_W$, but
it varies in a random way over larger distances in order to respect
causality.
The vector $V_i$ is also random, of course.
Its variation is due to the fact that the field $\phi$ makes a
random walk on the vacuum manifold of $\phi$.
The problem then is to estimate the field $F_{ij}$ over a length scale
$\sim N/m_W$.
If $N = 1$, then it follows that on dimensional grounds $F_{ij} \sim
m_W^2 \sim 10^{24} $ G, with probably an uncertainty of $\pm 1$ in
the exponent \cite{vach}.
For $N$ large, one should use a statistical argument.
In \cite{vach} it was argued that the gradients are of order
$1/ \sqrt{N}$, since $\phi$ makes a random walk on the vacuum manifold
with $\Del \phi \sim \sqrt{N}$, and since $\Del x \sim N$.
Thus $V_i$ is, in a root mean square sense, of the order
$1/ \sqrt{N}$, and hence $F_{ij}$ is of order $1/N$.
Taking further into account that the flux in a co-moving circular
contour is constant, the field must decrease like $1/ a(t)^2$, where
$a(t)$ is the scale factor \cite{vach}.
Using the fact that in the early universe $a$ goes like the
inverse temperature, the field
was then estimated to behave like
\beq
\langle F_{ij}\rangle_T~ \sim \frac{T^2}{N}                 \label{3}
\eeq
when the temperature of the universe is $T$.
For a scale of 100 kpc this leads to $\langle F_{ij}\rangle_{now} ~\sim
10^{-30} $ G, which is far too small
to explain the galactic fields (unless there exists some large scale
amplification mechanism).

In the present paper we shall present a different statistical scenario
where the gradient vectors are taken to be the basic stochastic variables.
Our considerations have been influenced by the fact that in large scale
3-dimensional computer simulations of the dynamo effect \cite{brand}
the computer uses a random initial magnetic field configuration.
We find that the very interesting expression \rf{2} obtained in
\cite{vach} can be interpreted statistically in such a way that the
mean magnetic field satisfies
\beq
\langle F_{ij}\rangle_T ~= 0~~,~~~~~\sqrt{\langle F^2_{ij}\rangle_T}~ \sim
\frac{T^2}{\sqrt{N}}~~.                         \label{4}
\eeq
Comparing the root-mean-square value \rf{4} with Eq. \rf{3},
we observe that the scaling behavior is weaker by $\sqrt{N}$.
This means that for a scale of 100 kpc
\beq
\sqrt{\langle F^2_{ij}\rangle_{today}} \sim  10^{-18} {\rm G}~~,   \label{5}
\eeq
which is very close (if not equal) to the value desired for the dynamo
effect.

It should be emphasized that in estimates it is reasonable, from the
point of view of the dynamo effect, to calculate the {\em rms} value
(or, more precisely, to estimate the {\em rms} value of the
projection of the random field on the dynamo eigenfunctions)
\cite{shuk}, whereas in actual computer simulations the field configuration
should be generated by the computer subject to the two
conditions \rf{4}.

\section{Discussion of the averaging procedure}

Let us now turn to the detailed arguments.
We wish to consider random fields walking around in space in a certain
number of steps.
Thus we replace the continuum by a lattice, where the points are
denoted by greek letters $\al, \ldots$.
We want to estimate the magnetic field over a {\em linear} scale
(which at most is equal to the horizon scale).
Thus, we consider a curve consisting of $N$ steps in the
lattice, and we define the mean value
\beq
\oB = \frac{1}{N} \sum^N_{i=1} B^{\al_{i}}~~,          \label{6}
\eeq
where $B$ is a component of the magnetic field, and where the
lattice points $\al_i$ are on the curve.

Now this curve is arbitrary, and we could take any other curve.
We therefore define the average $\langle  \ldots \rangle$, which averages over
curves spanning an $N^3$ lattice, i.e. over all space (this is well defined on
a lattice space; one could e.g. take the set of all curves that are parallel to
one of the sides of the $N^3$ lattice).
Then, for example,
\beq
\langle  \oB \rangle = \frac{1}{N} \langle \sum^N_{i=1} B^{\al_{i}}
\rangle~~,                               \label{7}
\eeq
which means that for each curve with $N$ steps the mean value
$\oB$ is computed, and this is done for a set of curves which span an
$N^3$-lattice,
and the average is then computed.
Therefore $\langle \oB\rangle$ depends in general on $N$, but for simplicity of
notation we shall leave out the explicit reference to this dependence.
We wish to emphasize that the ensemble average \rf{7} takes into account
the field value at each lattice point, so that {\it the average is really over
the whole lattice volume}.

Similarly, one can define higher moments such as
\beq
\langle  \oB^2 \rangle = \frac{1}{N^2}
\sum^N_{i,j=1} \langle B^{\al_{i}} B^{\al_{j}}\rangle~~,  \label{8}
\eeq
together with quantities like
\bdm
\langle ( \oB - \langle \oB\rangle )^2  \rangle~~.
\edm
Note that in  \rf{8}  the sum is over
curves of length $N$ steps of the non-local quantity
$\langle B^{\al_{i}} B^{\al_{j}} \rangle$.

\section{Random Higgs gradients}

In this section we shall present our main statistical assumptions.
The general point of view is that when one has a random system it
is necessary to specify the statistical distribution
and also which variable is to be considered as a stochastic variable.
These two specifications are the necessary boundary conditions.

In \cite{vach} the stochastic variable was taken to be the Higgs field
itself which varies over the vacuum manifold.
However, it is clear that also the gradient vectors $V_i$ are
stochastic, and in our scenario we assume that these
vectors are the relevant stochastic variables.
This is because they directly specify whether there is a magnetic
field or not, whereas this is only true indirectly for the Higgs
field itself.
Also, the vectors $V_i$ are relevant for questions of alignment
between neighbouring domains.
Thus, a scenario which takes the gradient vectors as the basic
stochastic variables is rather natural.
In this scenario the vectors and the resulting magnetic field have
only short-range correlations.

We now return to the expression \rf{2} of the magnetic field in
terms of the Higgs gradients $V_i$.
It is convenient to split these fields into real and imaginary parts,
\beq
V_i ({\bfx}) = R_i ({\bfx}) + i I_i ({\bfx})~~,                \label{9}
\eeq
where $R_i$ and $I_i$ are real vectors.
We consider the system at a fixed time.
The cosmological boundary condition is then that $R_i$ and $I_i$ are
random fields.
We now make the following assumptions:
\begin{itemize}
\item[(i)] The random fields have a Gaussian distribution.
Thus, the mean value of some quantity $Q$ is given by
\beq
\langle Q\rangle = \prod_{\al,i} \int \frac{d^3 R^\al_i}{D} \frac{d^3
I^\al_i}{D}
Q~e^{- \lam (R^\al_i - \langle  \oR_i \rangle)^2 -
\lam(I^\al_i - \langle  \oI_i\rangle )^2}~~,
                                          \label{10}
\eeq
where $D$ is a normalization factor defined such that $\langle 1\rangle =1$,
and $\lam$ is a measure of the inverse
width.
The quantities $\oR_i$ and $\oI_i$ are the mean values of $R_i$ and $I_i$
defined along a curve of length $N$ steps.\footnote{Note that this implies
that the mean value in a point is assumed to be equal to the mean value
computed along all curves of length $N$. Thus the mean values can
depend on $N$.}
Thus, eq. \rf{10} is relevant for a 3-dimensional world which is an
$N^3$ lattice.

\item[(ii)] We assume that the mean values are isotropic, i.e.
$\langle  \oR_1\rangle = \langle  \oR_2\rangle = \langle  \oR_3\rangle$ and
$\langle  \oI_1\rangle = \langle  \oI_2\rangle = \langle  \oI_3\rangle $.
\end{itemize}

Assumption (i) is certainly the simplest way of implementing lack of
correlation of the gradient vectors over distances compatible
with the horizon scale, whereas
assumption (ii) is natural as there is no reason to expect any
preferred direction.

It should be noted that the distribution \rf{10} factorizes into an
$I$-part and an $R$-part.
Thus, for any expectation value consisting of $I$'s and $R$'s one has
factorization,
\beq
\langle \oR^{\al_{1}}_{i_{1}} \ldots \oR^{\al_{n}}_{i_{n}}
{}~~\oI^{\beta_{1}}_{j_{1}} \ldots  \oI^{\beta_{m}}_{j_{m}} \rangle
= \langle \oR^{\al_{1}}_{i_{1}} \ldots
\oR^{\al_{n}}_{i_{n}}\rangle \langle \oI^{\beta_{1}}_{j_{1}}
\ldots \oI^{\beta_{m}}_{j_{m}}   \rangle~~.                     \label{11}
\eeq
This property turns out to be very useful in computing the higher moments.

\section{The expectation value of the magnetic field}

First we consider the expectation value of a component $B_i$ of the
magnetic field.
{}From the expression \rf{2} we find that
\beq
B_i  =  \frac{1}{2} \ve_{ijk} F_{jk} = - i~\ve_{ijk} V^\dagger_j V_k
 =  2 \ve_{ijk} R_j I_k~~.                    \label{12}
\eeq
Thus
\beq
\oB_j  =  \frac{1}{N} \sum^N_{i=1} B^{\al_{i}}_j
 =  2 \ve_{jlk} \frac{1}{N} \sum^N_{i=1} R^{\al_{i}}_l I^{\al_{i}}_k~~.
                                                \label{13}
\eeq
Hence
\bea
\langle \oB_j \rangle & = & \frac{2}{N} \ve_{jlk} \langle
\sum^N_{i=1}~R^{\al_{i}}_l
                I^{\al_{k}}_k \rangle \nn
         & = &  \frac{2}{N} \ve_{jlk} \langle \sum^N_{i=1}~
              (R^{\al_{i}}_l - \langle \oR_l \rangle )
                (I^{\al_{i}}_k - \langle \oI_k \rangle)
 + N \langle \oR_l \rangle \langle \oI_k \rangle \rangle~~.
                                    \label{14}
\eea
Now, due to the factorization \rf{11}, the first term on the right-hand
side of the last Eq. \rf{14} vanishes,\footnote{Because
$\langle R^{\al_{i}}_l - \langle  \oR_l \rangle\rangle = \langle I^{\al_{i}}_k
- \langle \oI_k \rangle\rangle = 0$
for symmetry reasons.}
and hence
\beq
\langle \oB_j \rangle = 2 \ve_{jlk} \langle \oR_l\rangle\langle \oI_k\rangle =
\ve_{jlk} \left(
\langle  \oR_l\rangle\langle \oI_k\rangle - \langle \oR_k\rangle\langle
\oI_l\rangle \right) = 0       \label{15}
\eeq
because of the isotropy assumption (ii).
Consequently {\em the mean value of the magnetic field vanishes},
as announced in the Introduction.

It should be noticed that if we did not assume isotropy, then
$ \langle  \oB_j\rangle \neq 0$ in general.
If we then use Vachaspati's argument \cite{vach}, $\langle \oR_i\rangle$ and
$\langle \oI_i\rangle$
behave like $1/ \sqrt{N}$, and hence from \rf{15} one would find
$\langle B_j\rangle \sim 1/N$.
Thus, our method of averaging over curves leads to the same result
as found previously, if we do not assume isotropy.
However, we believe that isotropy is a natural assumption.

\section{The root mean square of the magnetic field}

The second order moment is given by
\bea
\langle \oB^2_i \rangle & = & \frac{4}{N^2} \sum_{\al \beta} \langle R^\al
R^\beta \cdot
       I^\al I^\beta - R^\al I^\beta \cdot I^\al R^\beta \rangle \nn
& = &   \frac{4}{N^2}  \sum_{\al \beta} \left\{ \langle R^\al R^\beta \rangle
       \langle I^\al I^\beta\rangle - \langle R^\al_i R^\beta_j\rangle
\langle  I^\beta_i  I^\al_j \rangle \right\}~~,
                                          \label{16}
\eea
where we used the factorization \rf{11}.
Now
\bea
\langle R^\al_i R^\beta_j \rangle& =& \prod_\g \int~ \frac{d^3R^\g}{D}~R^\al_i
R^\beta_j
e^{- \lam (R^\g - \langle \oR\rangle)^2} \nn
&=& \frac{1}{2 \lam} \del_{ij} \del^{\al \beta} + \prod_\g \int
\frac{d^3R^\g}{D} \left[ \langle \oR_i\rangle R^\beta_j +
\langle \oR_j\rangle R^\al_i
 - \langle \oR_i\rangle \langle \oR_j\rangle \right]
e^{- \lam (R^\g - \langle \oR \rangle)^2}~,                \label{17}
\eea
and similarly for $\langle I^\al_i I^\beta_j\rangle$.
Further we have e.g.
\bea
\prod_\g ~\int~ \frac{d^3 R^\g}{D}~ R^\beta_j e^{- \lam (R^\g - \langle
\oR\rangle)^2}
&=&  \prod_\g ~\int~ \frac{d^3 R^\g}{D}~ (R^\beta_j - \langle \oR_j\rangle)
e^{- \lam (R^\g - \langle \oR\rangle)^2} + \langle \oR_j\rangle \nn
&=& \langle \oR_j\rangle~~,                                        \label{18}
\eea
i.e., the mean value in a given arbitrary point $\beta$ on the lattice is
equal to the mean value computed over all curves.
Using \rf{17} and \rf{18} in \rf{16} we get
\beq
\langle \oB^2_i\rangle = \frac{4}{N^2} \sum_\al \left( \frac{3}{2 \lam^2} +
\frac{1}{\lam}
( \langle \oI\rangle^2 + \langle \oR\rangle^2) \right)
 + \frac{4}{N^2} \sum_{\al \beta}
\left( \langle \oR\rangle^2  \langle \oI\rangle^2 -
(\langle  \oR\rangle \langle \oI\rangle)^2 \right)~~.         \label{19}
\eeq
The first term is $\cO (\frac{N}{N^2}) = \cO (\frac{1}{N})$.
The last term, being the square of the mean value, actually vanishes
because of isotropy: If we take
\beq
\langle \oR\rangle = \frac{1}{\sqrt{3}} (r,r,r)~~;~~
\langle \oI\rangle = \frac{1}{\sqrt{3}}
(c,c,c)~~;~~ \langle \oR\rangle^2 = r^2~~;~~\langle \oI\rangle^2 = c^2~~,
         \label{20}
\eeq
then
\beq
\langle \oR\rangle^2 \langle \oI\rangle^2 -
(\langle \oR\rangle \langle \oI\rangle)^2 = r^2c^2 -
\left(\frac{1}{3}(3rc)\right)^2 = 0~~.                  \label{21}
\eeq
Thus we conclude that the {\em rms} value of the magnetic field is given by
\beq
\sqrt{\langle \oB^2_i\rangle} = \frac{2}{N} \sqrt{\sum_\al \left( \frac{3}{2
\lam^2}
+ \frac{1}{\lam} ( \langle \oI\rangle^2 + \langle
\oR\rangle^2 ) \right)} \sim
\cO (\frac{1}{\sqrt{N}}) ~~,                            \label{22}
\eeq
as announced in the Introduction.

The reason for this slow decrease is the fact that isotropy prevents the
mean value from entering in $\langle \oB_i\rangle$ and $\langle \oB^2\rangle$,
and that the correlations
of the gradient vectors are of short range.

\section{Another probability distribution}

It should be mentioned that the scenario developed in the previous
section is by no means unique from the point of view of producing
short range correlations.
Instead of assuming that the vector $V_i$ has a random distribution
one could make the assumption that  the magnetic field
$B_i$ itself has a random distribution, with the probability
\beq
\prod^N_{i=1} \prod^3_{j=1} e^{- \lam(B^{\al_{i}}_j)^2}
\frac{d^3B^{\al_{i}}}{D}~~,
                                                      \label{23}
\eeq
where $D$ is a normalization factor.
Here we have assumed that $\langle \oB_i\rangle = 0$.
Then one has
\beq
\langle \oB^\al_i \oB^\beta_j\rangle =
\frac{1}{2 \lam} \del_{ij} \del^{\al \beta}
                                                \label{24}
\eeq
and hence
\bea
\langle \oB^2\rangle& = &
\frac{1}{N^2}~ \sum^N_{i,j=1} \langle  B^{\al_{i}} B^{\al_{j}} \rangle  \nn
& = & \frac{1}{N^2}~ \sum^N_{i=1} \langle ( B^{\al_{i}})^2\rangle
\sim \cO(\frac{1}{N})~~~~,
                                                   \label{25}
\eea
i.e. the same as the previous result.
It should be noted that this result appears in spite of the fact that
the distribution \rf{23} is very different from the distribution
\rf{11} of the  vector field, since e.g. \rf{23} contains
correlations between the $I_i$ and $R_i$ fields. Also, it should
be noticed that the reason for the result \rf{25} is that the
magnetic field has only short-range correlations.

The distribution \rf{23} is the one which is usually assumed in
solid state physics when dealing with a random magnetic field.
In the continuum version it reads
\beq
e^{- \lam \int d^3x B_i(x)^2} {\cD} B(x)          \label{26}
\eeq
and one then has
\beq
\langle B_i ({\bfx}) B_j ({\bfy})\rangle =\frac{1}{2
\lam} \del_{ij} \del^3({\bfx} - {\bfy})~~~~,
                                                \label{27}
\eeq
where the $\del$-function is assumed to be smeared.

Let us  also  comment on the case when $\langle \oB\rangle \neq 0$.
For the fluctuations we would have
\beq
\langle \oB^2\rangle = \frac{1}{N^2} \sum_{\al \beta} \langle B^\al
B^\beta\rangle
 =  \langle \oB\rangle^2 +
\frac{1}{N^2} \sum_{\al \beta} \langle(B^\al - \langle \oB\rangle )
(B^\beta - \langle \oB\rangle) \rangle ~~~~.                         \label{28}
\eeq
With only short range correlations, i.e. with
\beq
\langle (B^\al - \langle \oB\rangle)(B^\beta - \langle \oB\rangle) \rangle =
\del^{\al \beta} \langle (B^\al - \langle \oB\rangle)^2 \rangle     \label{29}
\eeq
this gives
\beq
\langle \oB^2\rangle = \langle \oB\rangle^2 + \cO (\frac{1}{N})~~~~.\label{30}
\eeq
Thus, if one has $\langle \oB\rangle \sim \cO (1/N)$ as in ref. \cite{vach},
and if the correlations are of short range, then the dominant
term is the fluctuations $\cO (1/N)$ in Eq. \rf{30}.
Therefore, even in this case one has $ \langle \oB\rangle^2 \ll \langle
\oB^2\rangle$,
and hence the field should be estimated from the {\em rms}-value
$\sqrt{\langle \oB^2\rangle}$ when $N$ is large, not from $\langle \oB\rangle$,
and the {\em rms}-value is again effectively of order
$1/ \sqrt{N}$.
\section{The energy-momentum tensor}
We should briefly discuss the consequences of the results described in Sections
4-6 for the energy-momentum tensor $T_{\mu\nu}$. After a calculation analogous
to that in Eqs. (16)-(22) we obtain
\beq
\langle \oB_i\oB_l\rangle = \frac 4N\delta_{il}\left[\frac 1{2\lambda^2}+\frac
1{2\lambda}\left(\langle\oI^2\rangle+\langle\oR^2\rangle\right)\right ]-\frac
2{N\lambda}\left(\langle\oI_i\rangle
\langle\oI_l\rangle+
\langle\oR_i\rangle
\langle\oR_l\rangle\right)~~.           \label{36}
\eeq
Contraction over $i$ and $l$ of course reproduces Eq. \rf{22}. We see that
the expectation value in Eq. \rf{36} has two terms, an isotropic and an
isotropic term. This is also carried over to the energy-momentum tensor
\beq
T_{00}=\frac 12 B^2,~~T_{il}=\frac 12\delta_{il}B^2-B_iB_l~~.  \label{37}
\eeq
Consider first the case $\langle\oI_l\rangle =
\langle\oR_l\rangle =0$. Then
\beq
\langle T_{il}\rangle=\frac 12\delta_{il}\langle B^2\rangle -\langle
B_iB_l\rangle
=\frac 16\delta_{il}\langle B^2\rangle~~,         \label{38}
\eeq
so that the equation of state is still isotropic: $p=\frac 13\rho$. However,
in the case where  $\langle\oI_l\rangle$ and
$\langle\oR_l\rangle$ are non-vanishing we obtain instead of Eq. \rf{37}
\beq
\langle T_{il}\rangle= \frac 1{N\lambda^2}\delta_{il}+\frac
2{N\lambda}\left(\langle\oI_i\rangle
\langle\oI_l\rangle+
\langle\oR_i\rangle
\langle\oR_l\rangle\right)~~.           \label{39}
\eeq
Hence in this case $\langle T_{il}\rangle$ has an additional anisotropic
part.

If the field $B_i$ itself is a Gaussian random field, we find that
\bea
\langle T_{00}\rangle &=& \frac 12\langle (B-\langle\oB\rangle )^2\rangle+\frac
12\langle\oB\rangle^2~~,\nn
\langle T_{il}\rangle &=& \frac 12\delta_{il}~[\frac 13 \langle
(B-\oB)^2\rangle+\langle\oB\rangle^2]-\langle\oB_i\rangle\langle\oB_l\rangle~~,
\label{40}
\eea
so in general there also appears an anisotropic part in $\langle
T_{il}\rangle$.
If $\langle\oB\rangle$ is ${\cal O}(1/N)$ as in \cite{vach} the anisotropic
part
is subdominant because of Eqs. \rf{28} - \rf{30}. Thus, to the leading order
$1/N$\ $\langle T_{il}\rangle$ is isotropic.

In principle the magnetic energy-momentum tensor discussed above contributes to
the expansion of the universe. However, in the present universe its effect is
extremely small and can safely be ignored. In the early universe this may not
be so. Our results show that in principle there may exist non-isotropic
stresses of magnetic origin, which in a statistical sense could contribute to
produce
turbulence in the primeval plasma, giving possibly rise to an early "universal"
dynamo effect. Magnetically generated  plasma flows might also be important
e.g.  for the dynamics of the QCD phase transition.
\section{Consequences of the magnetic field}
\def\brms{B_{\rm rms}}
\def\rhob{\rho_B}
\def\rhog{\rho_\gamma}
We assume that at the electroweak scale $T=T_0\simeq 100$ GeV the coherence
length of the {\em rms} field is $\xi_0 \simeq 1/T_0$ so that in terms of the
physical distance $L$, we have $N=L/\xi_0$. The magnetic field is frozen
at that time, so that at later times the original coherence
length is redshifted by the expansion according to
\begin{equation}
\xi(t)={a(t)\over a_0}\xi_0~~~~. \label{31}
\end{equation}
The frozen--in magnetic field is also redshifted by the expansion of the
universe. Thus
at later times at the distance scale $L$,
\begin{equation}
\brms (t,L) =B_0\left( {a_0\over a(t)}\right)^2{1\over\sqrt{N}}=B_0
\left({t_0\over t_*}\right)^{3\over 4}\left({t_*\over
t}\right)\left({\xi_0\over L}\right)^{1\over 2},            \label{32}
\end{equation}
where $T_0^2t_0=0.301\; M_P/\sqrt{g_*(T_0)}$ with $g_*$ the effective number
of degrees of freedom, and $t_*\simeq 1.4\times 10^3 (\Omega_0h^2)^{-2}$ yrs is
the time when the universe becomes matter dominated; for definiteness, we
shall adopt the the value $\Omega_0h^2=0.4$, which is the upper limit
allowed by the age of the universe. We shall also assume that $B_0\simeq
10^{24}$ G.

We may easily find from Eq. \rf{32} the size of the cosmological field today,
which could have acted as the seed field for the dynamo mechanism.
Taking $t=1.5\times 10^{10}$ yrs and $L=100$ kpc (corresponding to $N=1.0\times
10^{24}$), we find that today the
cosmological field at the scale of intergalactic distances is
\begin{equation}
\brms =4\times 10^{-19}\ {\rm G}~~.       \label{33}
\end{equation}
This seems to be exactly what is required for the numerical dynamo simulations
to produce the observed galactic magnetic fields of the order $10^{-6}$ G. The
inherent uncertainties in the estimate \rf{33} are: the value of $\Omega_0h^2$
used for computing $t_*$; the time at which the magnetic field froze, or $T_0$;
the actual value of the field $B_0$. Therefore one should view \rf{33} as an
order--of--magnitude estimate only.

We should also check what other possible cosmological consequences the
existence of the random magnetic field, Eq. \rf{32}, may have. Let us first
note that the energy density $\rhob$ in the {\em rms} field is very small.
In the
radiation dominated era we find that the energy density within a horizon volume
$V$ is
\begin{equation}
\rhob ={1\over 2V}\int_0^{r_H} d^3{\bf r}\brms^2={3\over 4}
B_0^2\left({T\over T_0}\right )^4{1\over r_HT}~~.     \label{34}
\end{equation}
The
horizon distance is $r_H=2t$ so that $\rhob \sim T^5/M_P\ll \rhog$, and
the magnetic field contribution to the total energy density is negligible.

In principle, magnetic fields could modify primordial nucleosynthesis. For
instance, it
has been argued \cite{hector} that protons actually become heavier than
neutrons in a large enough magnetic field. In our case, however, the magnetic
field is glued to the charges according to Eq. \rf{1}
so that the relative velocity is zero.
Note that this
prevents the charged particles in the plasma from accelerating by emitting
synchrotron radiation, as would happen if there were a constant background
field. At the onset of nucleosynthesis, at about $T\simeq 1$ MeV, the effect
of the {\em rms} field on the weak reaction rates which change protons to
neutrons and determine the crucial $n/p$--ratio turns also out to be
negligible. Since at
$T\simeq 1$ MeV
$n\leftrightarrow p$ reactions have scattering lengths of the order of the
horizon length, from Eq. \rf{32} we find that the relevant field at that scale
is only
$\brms\simeq 1500$ G. Also, creating a thermal population of right--handed
neutrinos, disastrous for the successful prediction of primordial light element
abundances, via scattering of left--handed neutrinos off the magnetic field
\cite{eos}
may not be possible in our case because the mean squared length of the field
fluctuation is expected to be very short, $\cO (1/T)$. This means
that the $\nu_L\leftrightarrow\nu_R$ transition probability would be
very much suppressed \cite{es}.
However, this issue can only be settled by detailed dynamical
considerations, for example by computer simulations.


Perhaps more interesting is the role of the {\em rms} field at the QCD phase
transition \cite{qcd}. It is believed to be of first order, but the details
of bubble nucleation depend on the largely unknown dynamical details of
QCD. In simplistic nucleation theory one obtains, by comparing the nucleation
rate with the Hubble rate, for the size of the critical
bubble about 10 fm. The distance $d$ between nucleation centers depends on
the amount of supercooling, but with reasonable assumptions $d\gtwid 10^{-2}$
m.
This is then the scale at which the bubbles of new phase will feel the random
background field. We find
\beq
\brms (t_{\rm QCD},d)\ltwid 1.8\times 10^9\ {\rm G}~~.
              \label{35}
\eeq
Whether this  has an effect on the QCD phase transition or not depends
to a large extent on whether the quarks are glued to the flux lines.
We shall not discuss this issue further.

\section{Conclusions}

We have shown that if the derivative of the (logarithm of the) Higgs
field is a random variable, then Vachaspati's construction \cite{vach}
leads to a magnetic field in the present day universe which has the
right order of magnitude from the point of view of the galactic dynamo
mechanism
\cite{dynamo}.

It should also be mentioned that this result is more general, as is
clear from Sect. 6.
This is because if by some mechanism a Gaussian  random magnetic field is
generated at the electroweak phase transition (where it is always of
order $m_W^2$ over a correlation   length on dimensional grounds),
then Eq. \rf{25} shows that the {\em rms}-value behaves like
$1/ \sqrt{N}$, if $\langle \oB \rangle = 0$.
However, even if $\langle \oB \rangle \neq 0$, it follows from
Eq. \rf{30} that the {\em rms} value is at least of order
$1/ \sqrt{N}$.
This is because either $\langle \oB \rangle^2$ is larger than or
equal to $1/N$, in which case the {\em rms} value is larger than
or equal to $1/ \sqrt{N}$ or $\langle \oB \rangle^2$ is
less than $1/N$, in which case the fluctuation term of order $1/N$
dominates, and the {\em rms}-value is of order $1/ \sqrt{N}$.
Thus, if there exists any mechanism which produces a Gaussian
random magnetic field at the electroweak phase transition, it will
produce a result which is larger than or equal to the field required
by the dynamo effect. It therefore appears that there exists a good case
for the primordial origin of the observed galactic magnetic fields.

If we had considered a magnetic field performing a random walk in a given
volume, or a random magnetic flux through a given surface, we would have
obtained
the scalings $1/N^{3/2}$ and $1/N$, respectively. Then one still would have had
to weight these spatial averages by the statistical distribution, but this
would then have induced double counting of the lattice points. Therefore
we have considered random walk only along a given curve, together with an
ensemble average over all curves spanning a given volume.
Therefore our method takes into account the field value at each lattice point,
and in this sense the ensemble average is really over the whole lattice volume.
Only a detailed dynamical
simulation can however tell for sure whether our seed field
actually gives rise to the observed galactic magnetic field. A definite
prediction of our scenario is that
the seed field has $\langle \oB\rangle =0$ by virtue of the averaging
procedure, and that only the {\em rms} field is non-vanishing.
\vskip 1truecm\noindent
{\Large\bf Acknowledgements}
\vskip 0.5truecm\noindent
We are grateful for A. Brandenburg and A. Shukurov for explaining the
galactic dynamo, V. Semikoz for information on magnetohydrodynamics,
and T. Vachaspati and
A. Vilenkin for useful discussions.

\newpage
\def\yana#1#2#3{Astron.\ Astrophys.\ {\bf #2}\ {(19#1)}\ {#3}}
\def\yanas#1#2#3{Astron.\ Astrophys.\ Suppl.\ {\bf #2}\ {(19#1)}\ {#3}}
\def\yapj#1#2#3{Astrophys.\ J.\ {\bf #2}\ {(19#1)}\ {#3}}

\end{document}